# Symmetry Protected Josephson Supercurrents in Three-Dimensional Topological Insulators

S. Cho[1], B. Dellabetta[2], A. Yang[3], J. Schneeloch[3], Z. J. Xu[3], T. Valla[3], G. Gu[3], M.J. Gilbert*[2], N. Mason**[1]

**Coupling the surface state of a topological insulator (TI) to an s-wave superconductor is predicted to produce the long-sought Majorana quasiparticle excitations[1-4]. However, superconductivity has not been measured in surface states when the bulk charge carriers are fully depleted, i.e., in the true topological regime that is relevant for investigating Majorana modes[5, 6]. Here, we report measurements of DC Josephson effects in TI-superconductor junctions as the chemical potential is moved from the bulk bands into the band gap, or through the true topological regime characterized by the presence of only surface currents. We examine the relative behavior of the system at different bulk/surface ratios, determining the effects of strong bulk/surface mixing, disorder, and magnetic field. We compare our results to 3D quantum transport simulations to conclude that the supercurrent is largely carried by surface states, due to the inherent topology of the bands, and that it is robust against disorder.**

Three-dimensional topological insulators, such as $Bi_2Se_3$ and $Bi_2Te_3$, are characterized by the existence of a bulk insulating energy gap, gapless Dirac surface states, and a $\pi$-Berry's phase[7]. For 3D TIs coupled to s-wave superconductors, the winding of the superconducting vortices can counteract the $\pi$-Berry's phase, resulting in zero-energy Majorana fermions produced at the interface[1]. Such Majorana fermions may be topologically protected from decoherence, and could play a significant role in solid state implementations of a quantum computer. A requisite step in the search for Majorana fermions is to understand the nature and origin of the supercurrent generated between superconducting contacts and a TI. Previous measurements of $Bi_2Se_3$- or $Bi_2Te_3$-superconductor junctions[5, 6, 8, 9] have demonstrated that the supercurrent can be tuned by a gate voltage[6] and exhibits Josephson effects such as Fraunhofer patterns[5, 9]; it has also been argued that the supercurrent is carried by surface states, though the mechanism was not well-understood[5]. Fundamental questions remain, particularly concerning the behavior through the Dirac point and the effects of disorder. We approach these by measuring the supercurrent over a wide-range of gate voltages (chemical potentials)—through a clear ambipolar transport regime—and by comparing our results to full 3D quantum transport calculations that can include disorder. We find that the supercurrent is largely carried by surface states because of protected crystal symmetries, and is only suppressed when bulk/surface mixing is strong or at very low carrier densities. We further find that the supercurrent is not symmetric with respect to the conduction and valence bands, and that the Fraunhofer patterns are similar both within and outside of the topological regime.

[1] Department of Physics and Frederick Seitz Materials Research Laboratory, 104 South Goodwin Avenue, University of Illinois, Urbana, Illinois 61801, USA. **email: nadya@illinois.edu.
[2] Department of Electrical and Computer Engineering and Micro and Nanotechnology Laboratory, University of Illinois, 208 N. Wright St., Urbana, IL 61801, USA. *email: matthewg@illinois.edu
[3] Condensed Matter Physics and Materials Science Department, Brookhaven National Laboratory, Upton, NY 11973, USA

The major obstacles in reaching the topological regime in transport experiments are high n-doping in as-grown TI materials and the further increase of n-doping during mechanical cleavage of crystals[10-13]. We implemented three strategies to reduce doping in thin exfoliated films, utilizing: (1) Sb-doped TI materials, (2) chemical doping of the top surface, and (3) a backgate. It has been previously reported that growing 3D TI materials with *in-situ* Ca or Sb dopants reduces n-doping[10, 13-15]. Thus, we have grown $(Bi_xSb_{2-x})Se_3$ crystals having x=1.33 (see Methods). Angle resolved photoemission spectroscopy (ARPES) confirms the topological spectra of the crystals (Figure 2d). From Fig. 2d, it is evident that the bulk conduction band is located at an energy $E \sim 0.20$ eV above the Dirac point and gapless linear surface states exist inside the bulk band gap. The Fermi velocity calculated from the linear dispersion of the surface states is $\sim 4.5 \times 10^5$ m/s and the total surface carrier density at the bottom of the bulk band is estimated to be $k^2/4\pi \times 2$ (two surfaces) $= 1/2\pi(E/\hbar v_F)^2 \sim 0.8 \times 10^{13}/cm^2$.

Devices consisted of mechanically exfoliated $\sim 10$ nm thick films on $Si/SiO_2$ substrates[10-12] (which are used as backgates) having Ti(2.5 nm)/Al(140 nm) contacts (see Methods). We deposited the chemical dopant 2,3,5,6-tetrafluoro-7,7,8,8-tetracyanoquinodimethane (F4-TCNQ) on some devices, as shown schematically in Fig. 1a. F4-TCNQ has a strong electron affinity, and has been shown to effectively remove excess n-doping in $Bi_2Se_3$[12]. A backgate voltage was then used to fully deplete the bulk charge carriers and tune the chemical potential through the Dirac point. It has been shown that once the bulk charge carriers are mostly depleted, the Fermi levels of top and bottom surface states lock to each other and shift simultaneously, even with a single gate[12]. In this paper, we show experimental results from two devices, where "device 1" has chemical doping and "device 2" does not. It was possible to use a backgate to deplete bulk carriers in device 2 because of the low n-doping due to Sb; however, in this case a much larger backgate voltage was required to reach the Dirac point. All transport measurements were performed in a dilution refrigerator at the base temperature of $T = 16$ mK. Each device was configured with both a Hall bar geometry and with closely-spaced electrodes across which Josephson currents were measured, as shown in Figs. 1a and 1b.

Figure 2 shows resistivity and Hall measurements for device 1 at $T = 16$mK and with $B_{perpendicular} = 35$mT (to suppress superconductivity in the aluminum leads). We observe a peak in resistivity $\rho_{xx}$ (minimum in conductivity $\sigma_{xx}$) and a corresponding sign change in Hall carrier density $n_H$ near $V_g \sim -55$V, which signal that the charge carriers change from electrons to holes as the chemical potential passes through the Dirac point. By comparing the total surface carrier density at the bottom of the bulk band calculated from ARPES measurements, $n_H \sim 0.8 \times 10^{13}/cm^2$, to the $n_H$ vs. $V_g$ measured in Fig. 2b, we find that the bottom of the bulk band occurs near $V_g \sim -18$V. This indicates that $V_g < -18$V is the topological regime, where only surface states are occupied, while $V_g > -18$V is the regime where both surface and bulk states are populated.

We now turn to measurements of Josephson effects. Figure 1c shows *I-V* curves measured for device 2 at three different temperatures, for $V_g=0$. The *I-V* curves exhibit zero-voltage regimes for currents less than the critical current $I_c$. Above $I_c$, a finite voltage is measured as the sample transitions to the normal regime. The critical current can also be modulated by $V_g$: in Figures 3b and 3d, we show 2D plots of differential resistance *dV/dI* vs. current *I* vs. $V_g$, for devices 1 and 2, respectively. The purple regions where *dV/dI = 0* indicate superconducting regions, and the boundary corresponds to $I_c$ for a given $V_g$. Figures 3a and 3c show the corresponding normal state $dV/dI$ vs. $V_g$ for the two-terminal Josephson configuration (taken at B = 35 mT). For device 1, the Dirac point is identified with the maximum resistance peak at $V_g \sim -55V$, consistent with the Hall measurement shown in Figure 2; the Dirac point for device 2 is near $V_g \sim -140V$.

It is now possible to compare the behavior of the supercurrent in different transport regimes. Figure 3 shows that in both devices, the critical current decreases, non-monotonically, as the Dirac point is approached from the conduction band, and does not increase significantly in the valence band. Focusing first on device 1 (Fig. 3b), we note that the system does not seem to change behavior at the position of the bottom of the conduction band (BCB, $V_g \sim -18V$). Below the BCB, $I_c$ continues to decrease until it reaches the Dirac point, near which it saturates to a small but finite value. The finite $I_c$ near the Dirac point may be caused by residual densities in electron-hole puddles due to charged impurity potentials[12, 16-18]. Surprisingly, the critical current does not increase in the hole region ($V_g < -55V$); this may be related to asymmetric contact resistances[19] or to the lack of clear surface states in the valence band (as seen in Fig. 2d). In device 2 (Fig. 3d), $I_c$ becomes zero at large negative gate voltages, even before the Dirac point is reached. In fact, $dV/dI$ exhibits a peak for $V_g < -100V$, the height of which increases with lower temperatures. The different behavior of device 2 at low densities may be due to increased disorder or longer channel length[20].

In order to explain the unique features observed in $I_c$ vs. $V_G$, we compare the experimental data to transport simulations of a model that couples superconducting contacts to a 3D tight-binding Dirac Hamiltonian with an inverted mass gap; this creates a topologically non-trivial Josephson junction which retains the geometry of the samples (see Supplemental information). Supercurrents and density of states (DOS) profiles are calculated in the non-equilibrium Green's function formalism[21]. Fig. 4a shows results for a geometry similar to device 1 overlayed with the measurements; there is excellent correspondence between simulations and experiment. To understand the $I_c$ vs. $V_g$ behavior, we plot the bulk and surface DOS for device 1 in Fig. 4b. We note that above the Dirac point, $I_c$ closely follows the surface DOS throughout the entire energy range, which explains the non-monotonic behavior of $I_c$. The fact that the supercurrent closely matches the surface DOS profile implies that the Josephson current is predominantly carried by the topological surface band, independent of bulk characteristics. Figure 4c shows simulations for a geometry similar to device 2 overlayed with the measurements; again, the simulations match the experiments. The supercurrent has a shape similar to that of device 1, but an insulating regime arises near the Dirac point. This may suggest that for longer junctions the channel becomes insulating above a critical resistance[20].

To further understand the surface-dominated supercurrent, we added δ-function impurities of increasing strength to bulk and/or surface regions in the long-channel model (2:1 length to width ratio). The effect of disorder localized in the bulk versus disorder in both surface and bulk can be seen in Fig. 4c and 4d, respectively. Bulk disorder has little effect on $I_c$ until it becomes large enough to hybridize the surface band with the bulk, i.e. when the impurity strength is comparable to the band gap energy. Disorder applied equally to bulk and surface states, in contrast, results in a continuous degradation of current as disorder is increased. These simulations strongly suggest that the large majority of supercurrent is carried along the surface of the TI, and that this occurs because of crystal symmetries, not because of difference in bulk and surface mobilities. In addition, the supercurrent only degrades if the surface is encumbered with elastic scattering centers, or if bulk disorder is so large that surface-bulk band hybridization occurs.

Finally, we examine a further hallmark of the Josephson effect, Fraunhofer behavior, inside and outside of the topological regime. Figure 5 shows a plot of measured $I_c$ as a function of magnetic field $B$ in device 1 at different gate voltages. The resulting Fraunhofer patterns, the single-slit interference-like dependence of $I_c$ on $B$ in a Josephson junction, demonstrate that the supercurrent through the junction originates from the Josephson effect. Figure 5 shows that while the amplitude of $I_c$ changes with $V_g$, the period of oscillation of $I_c$ with $B$ does not, for measurements both within and outside the topological regime.

## Methods

### Crystal growth

The crystals of $Bi_{1.33}Sb_{0.67}Se_3$ were grown by using a modified floating zone method in which the melting zone is Se-rich BiSbSe. The materials of high purity 99.9999% Bi, Sb and Se were pre-melted and loaded into a 10mm diameter quartz tube. The crystal growth velocity in the quartz tube is 0.5mm per hour.

### ARPES

The ARPES experiments were carried out at the National Synchrotron Light Source, using the VUV undulator beamline U13UB, which is based on a 3-m normal incidence monochromator. The photon energy used in the study was 20.5 eV. The electron analyzer was a Scienta SES-2002, which uses a two dimensional micro-channel plate as a detector that collects simultaneously a wide energy window and a wide angular window (~14°) of excited photoelectrons. The combined energy resolution was around 8 meV, while the angular resolution was better than ~ 0.15°, translating into a momentum resolution of ~ 0.005Å$^{-1}$ at 20.5 eV photon energy. Samples were mounted on a liquid He cryostat and cleaved *in-situ* and measured at ~15 K in the UHV chamber with the base pressure $3\times10^{-9}$ Pa.

### Device fabrication and measurement

We mechanically exfoliated thin films from bulk $(Bi_{1.33}Sb_{0.67})Se_3$ crystals on 300nm SiO$_2$/highly n-doped Si substrates by the "scotch tape method"[22]; films of thickness ~10 nm were found by optical microscope and atomic force microscopy[10-12]. Device 1 has a thickness of 14 nm and dimensions $L$ ~ 0.1 μm and $W$ ~ 1.4 μm; device 2 has a thickness of 12 nm and dimensions $L$ ~ 0.5 μm and $W$ ~ 2.5 μm. Subsequently, electron beam lithography was performed to define a Hall bar and closely spaced junctions in a device. Brief surface cleaning with ion milling and Ti(2.5nm)/Al(140nm) deposition were performed at a base pressure ~ 1x10$^{-9}$ *Torr* with sample substrates cooled to $T$ ~ 77K. Immediately after lift-off, device 2 was wire-bonded and cooled down in a commercial dilution refrigerator. About 8 nm of F4-TCNQ (Sigma-Aldrich) was deposited on device 1 before wire-bonding and cool-down. We observed ~2.7 times higher back-gate capacitance ($n = C_g V_g / e$ where $C_g$ ~ 30 nF/cm$^2$) in device 1 (with chemical dopants), compared to $C_g \sim 11$ nF/cm$^2$ in device 2 (with no chemical dopants). Similar values of enhanced back-gate capacitance at low temperature in thin Bi$_2$Se$_3$ devices doped with polymer electrolyte were reported previously,[12] however the origin is not well understood.

**Acknowledgements**

NM, SC, and MJG acknowledge support from the ONR under grant N0014-11-1-0728. MJG and BD acknowledge support from the AFOSR under grant FA9550-10-1-0459. Device fabrication was carried out in the MRL Central Facilities (partially supported by the DOE under DE-FG02-07ER46453 and DE-FG02-07ER46471). The work at BNL was supported by the US Department of Energy, Office of Basic Energy Sciences, under contract DE-AC02-98CH10886. SC acknowledges technical assistance from J. Ku and useful discussions with D. Kim and M. Fuhrer.

**Figure Captions**

Figure 1| **Measurement configuration and supercurrent characterization** (**a**) Device and measurement schematic and (**b**) SEM image of device 1. The blue dots in (**a**) represent F4-TCNQ molecules. The Hall measurement was performed in the Hall bar by applying current $I$ and measuring longitudinal ($V_{xx}$) and Hall ($V_H$) voltages, while the Josephson effect was measured in the closely spaced junction by applying current $I_J$ and measuring voltage $V_J$. (**c**) $I$-$V$ curves measured in device 2 at $V_g$=0 for four different temperatures, showing typical supercurrent behavior.

**Figure 2| Normal state resistivity, Hall, and ARPES characterization.** (**a**) Longitudinal resistivity $\rho_{xx}$, (**b**) Hall carrier density $n_H$, and (**c**) longitudinal conductivity $\sigma_{xx}$ for device 1as functions of backgate voltage. The maximum in $\rho_{xx}$ (min in $\sigma_{xx}$) and divergence of $n_H$ indicate the location of the Dirac point (DP) and the presence of both carrier types. (**d**) ARPES data of the bulk crystal, showing the expected Dirac cone band structure of the surface state. The bottom of the bulk conductance band (BCB) is labeled (see Text). It is clear from ARPES and Hall data that the system passes through a true topological regime of only surface states.

**Figure 3| Dependence of supercurrents on gate voltage.** Differential resistance ($dV/dI$) versus gate voltage in the normal regime (B = 35mT) for (**a**) device 1 and (**c**) device 2, showing locations of Dirac points in the Josephson configuration. Two-dimensional plots of $dV/dI$ versus gate voltage $V_g$ and current $I$ for (**b**) device 1 and (**d**) device 2. The purple regions show the extent of the supercurrent, where $dV/dI = 0$; the boundaries of these regions correspond to $I_c$.

**Figure 4| Comparisons of experiment to 3D simulations which include disorder.** (**a**) Dependence of critical current $I_c$ on gate voltage $V_g$ for a short channel junction (1:3 length to width ratio) (blue), overlayed with experimental results from device 1 (dashed black). (**b**) Surface (dashed blue) and bulk (solid blue) DOS for the model TI Hamiltonian. The bulk is gapped, with the first states appearing near -40 V. Although the peak in bulk DOS near $V_g$~-10 V is not compatible with the monotonicity of bulk DOS seen in the ARPES data, Fig. 4a shows the model captures the salient transport physics, as the bulk DOS profile does not play a significant role in transport within the junction. The effect of (**c**) bulk disorder and (**d**) surface and bulk disorder on $I_c$ in a long channel junction (2:1 length to width ratio), for δ-impurity strengths equal to 0, 0.2, 0.5 and 1.0 times the bulk band gap energy. The supercurrent only decreases appreciably when the bulk disorder potential is large enough to begin hybridizing bulk and surface bands. Including surface disorder causes the supercurrent to degrade immediately, signifying that supercurrent flows primarily through the surface band. Experimental results from device 2 are overlayed on both plots (dashed black).

**Figure 5| Fraunhofer data**. Critical supercurrent $I_c$ versus perpendicular magnetic field $B$ at $T = $ 16mK for device 1 at different gate voltages, showing clear Fraunhofer diffraction patterns. The minima occur at intervals of $\Phi/\Phi_0 \sim 0.23n$, where n is an integer; however, we plot $I_c$ vs $B$ rather than $\Phi/\Phi_0$ because the short junction area may be ill-defined due to flux focusing[23].

Figure 1

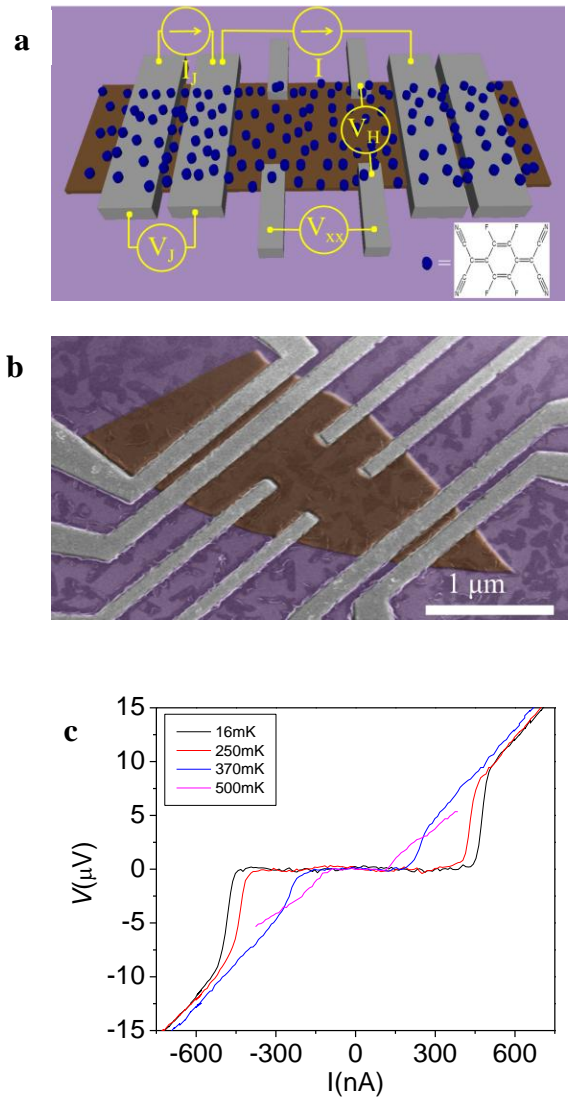

Figure 2

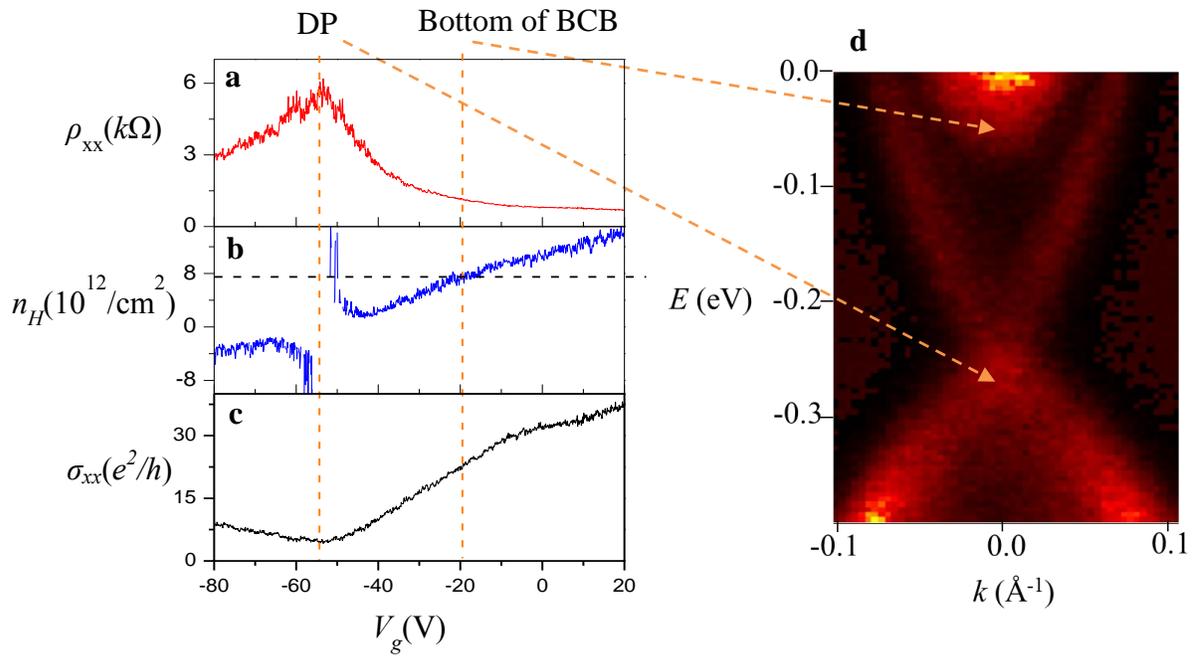

Figure 3

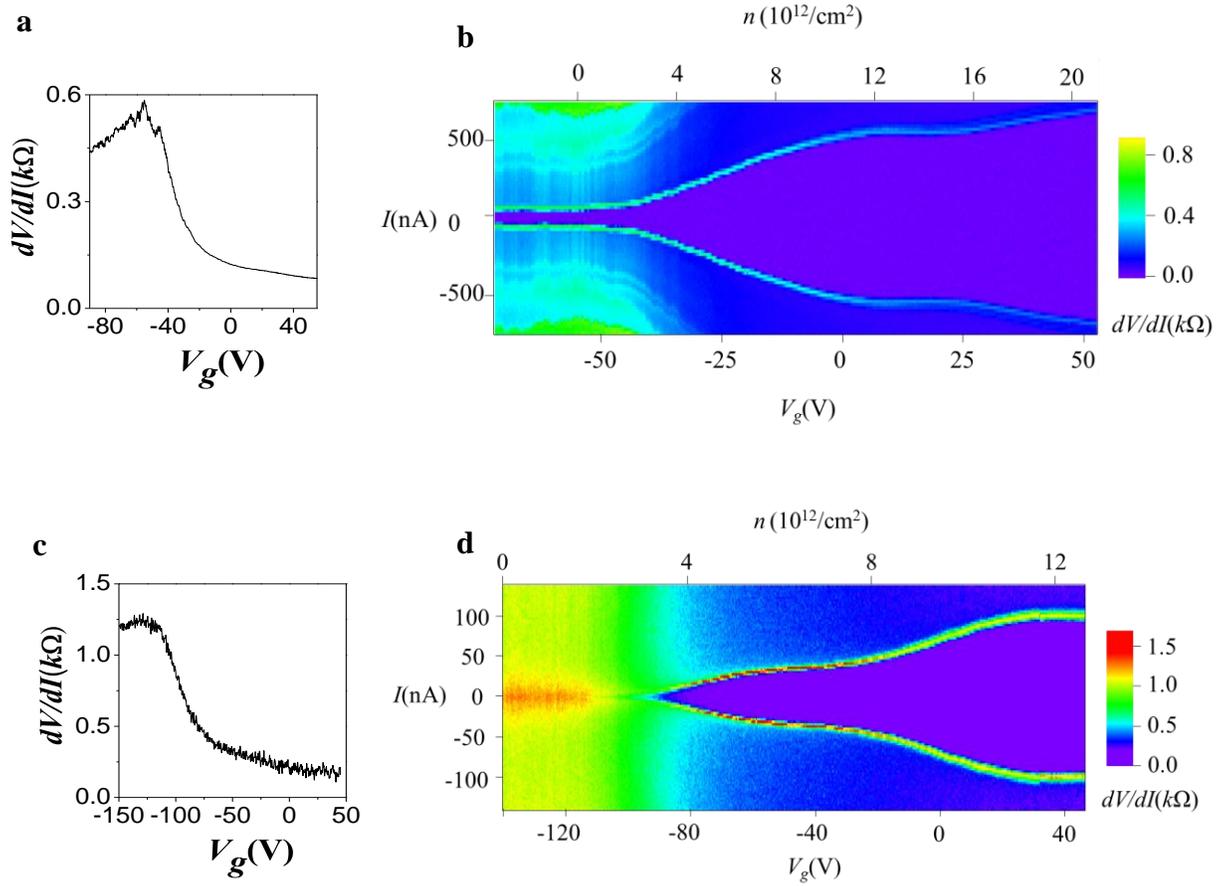

Figure 4

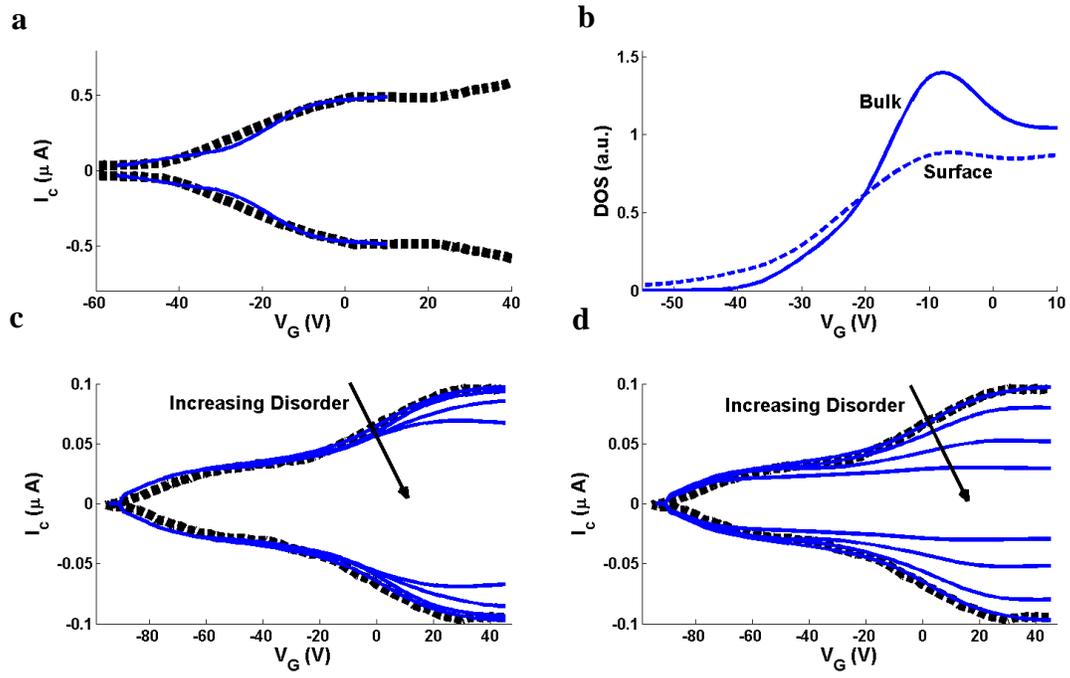

Figure 5

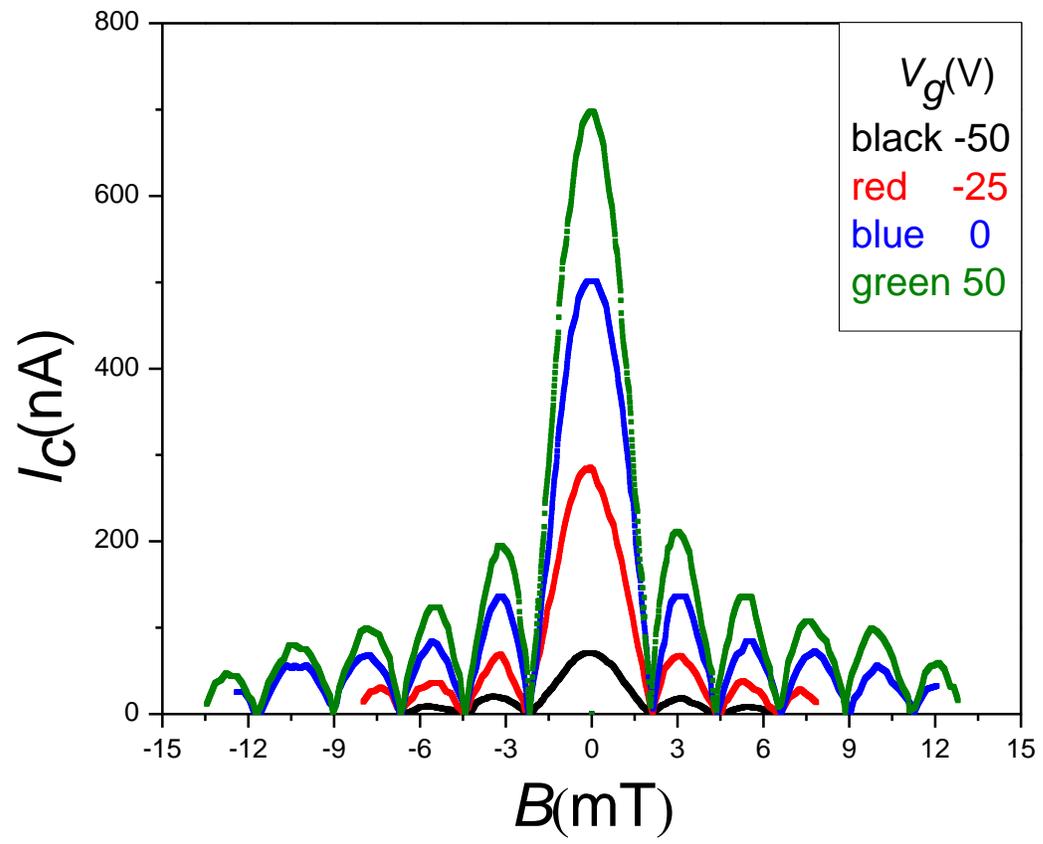

Supplementary Information

In this work, we consider an effective cubic-lattice model for the 3D Topological Insulator (TI) Hamiltonian with two bands (A, B) and two spin flavors ($\uparrow, \downarrow$). The resultant four-orbital Dirac Hamiltonian for the tight-binding model can be written as[1]

$$H_D = \sum_\mathbf{r} \left\{ (M\Gamma^0 - \mu\mathbb{I}_4)\Psi_\mathbf{r}^\dagger \Psi_\mathbf{r} + \sum_\delta \frac{B\Gamma^0 + iA\Gamma^\delta}{2} \Psi_\mathbf{r}^\dagger \Psi_{\mathbf{r}+\delta} \right\}, \quad (1)$$

where $\Psi_\mathbf{r} = (c_{A,\uparrow,\mathbf{r}} c_{A,\downarrow,\mathbf{r}} c_{B,\uparrow,\mathbf{r}} c_{B,\downarrow,\mathbf{r}})^T$ is the four-component spinor involving the spin and orbital degrees of freedom over all cubic lattice sites, $\mathbf{r} = (i,j,k)a_0$, which denotes lattice location with integers $i, j, k$ and lattice constant $a_0 = 1$Å. Only nearest neighbor hopping is considered, so that $\delta = \pm a_0 \hat{x}, \pm a_0 \hat{y}, \pm a_0 \hat{z}$. The Gamma matrices are definied as $\Gamma^\eta = \tau^x \otimes \sigma^\eta$ and $\Gamma^0 = \tau^z \otimes \mathbb{I}_2$. $\tau^\eta$ and $\sigma^\eta$ are the $2 \times 2$ Pauli matrices representing orbital and spin degrees of freedom, $\eta \in x, y, z$. $\mathbb{I}_N$ is the N × N identity matrix. The parameters $A$ and $B$ determine the band structure, and are set to $A = 1$ eV·Å and $B = 1$ eV·Å$^2$ for simplicity, so that the model is isotropic and preserves time reversal and particle-hole symmetry. The chemical potential $\mu$ is set by the gate voltage. The mass term $M = m - 3\frac{B}{a_0^2}$ yields a topologically nontrivial (trivial) phase when $m/B > 0$ ($m/B < 0$), and is set to $m = 1.5$ eV in the topologically nontrivial model so that bulk bands first appear near $\pm 0.5$ eV and the Dirac point lies exactly at 0.0 eV.

In order to introduce superconductivity, the four-orbital tight-binding model in Eq. 1 is expanded into a Bogoliubov-de Gennes (BdG) Hamiltonian as

$$\begin{aligned} H_{BdG} = & \sum_\mathbf{r} \Phi_\mathbf{r}^\dagger \begin{pmatrix} H_D(m) & \Delta(\mathbf{r}) \\ \Delta(\mathbf{r})^\dagger & -H_D^*(m) \end{pmatrix} \Phi_\mathbf{r} \\ + & \sum_\mathbf{r} \Phi_\mathbf{r}^\dagger \begin{pmatrix} H_D(\delta) & 0 \\ 0 & -H_D^*(\delta) \end{pmatrix} \Phi_{\mathbf{r}+\delta}, \end{aligned} \quad (2)$$

where $\Phi_r = (\Psi_\mathbf{r} \Psi_\mathbf{r}^\dagger)^T$ is an 8-component Nambu spinor and $H(m)$ and $H(\delta)$ denote the on-site and nearest neighbor hopping components in Eq. 1. In the mean-field description, $\Delta(\mathbf{r}) = \langle c_{\alpha,\downarrow,\mathbf{r}} c_{\beta,\uparrow,\mathbf{r}} \rangle$ is the order parameter which allows Cooper pairs to form between orbitals $\alpha$ and $\beta$ at a given lattice point as in a bulk s-wave superconductor (SC). We assume only intra-orbital pairing ($\alpha = \beta$) occurs, so that the $4 \times 4$ pairing matrix in Eq. 2 is

$$\Delta(\mathbf{r}) = \begin{pmatrix} 0 & \Delta_A(\mathbf{r}) & 0 & 0 \\ -\Delta_A(\mathbf{r}) & 0 & 0 & 0 \\ 0 & 0 & 0 & \Delta_B(\mathbf{r}) \\ 0 & 0 & -\Delta_B(\mathbf{r}) & 0 \end{pmatrix}. \quad (3)$$

The band-dependent order parameters $\Delta_A(\mathbf{r})$ and $\Delta_B(\mathbf{r})$ flip sign in Eq. 3 so that they obey the fermionic anticommutation relation. In order to simulate a physical Josephson Junction (JJ) of sufficient length, the proximity effect is assumed to occur only at the contact lattice points, decaying to zero into the TI channel. Simulations in which the proximity effect exponentially decays a few lattice points into the channel yield qualitatively similar results, with slightly higher currents due to higher coupling in the channel provided the superconducting wave functions of the contacts are separated by several lattice points.

The general expression for the retarded self-energy term of the left SC contact with order parameter $\Delta_L = |\Delta_L|e^{i\chi_L}$ is[2]

$$\Sigma_L^r(\omega) = \begin{pmatrix} \widetilde{m}_L(\omega + i\eta) & \widetilde{\Delta}_L(\omega + i\eta)e^{i\chi_L} \\ \widetilde{\Delta}_L(\omega + i\eta)e^{-i\chi_L} & \widetilde{m}_L(\omega + i\eta) \end{pmatrix}. \quad (5)$$

The effective on-site and pairing potentials for a given complex frequency in Eq. 5 are

$$\begin{aligned}
\widetilde{m}_L(z) &= -\frac{\Gamma}{2}\frac{z}{\sqrt{\Delta_L^2-z^2}}\\
\widetilde{\Delta}_L(z) &= \frac{\Gamma}{2}\frac{\Delta_L}{\sqrt{\Delta_L^2-z^2}},
\end{aligned} \qquad (6)$$

with the tunneling rate set to $\Gamma = 1$ for highly transparent interfaces. A similar expression for the right SC contact can be made as well. These matrices are only nonzero for lattice points at which the lead comes into contact with the channel. The left and right SC contacts are set with order parameter magnitude $|\Delta_L| = |\Delta_R| = 1.0$ meV, and the difference between the relative phases is a degree of freedom defined as $\chi_{LR} = \chi_L - \chi_R$.

This model provides a framework for studying the DC Josephson effect, where no voltage drop occurs between the left and right leads ($\mu_L = \mu_R = 0.0$ V) and supercurrent flow occurs only when there is a superconducting phase differential between the two contacts. The superconducting contacts in the simulations are set to a fixed phase differential, $\chi_{LR} = 0.8\pi$. A higher phase differential decreases current, as the confinement of bound states opens a gap in the bound state spectrum near $\chi_{LR} = \pi$, which causes supercurrent to vanish. This is again an artifact of small system size, however the transport dynamics of the DC Josephson effect remain qualitatively equivalent over the entire range of $\chi_{LR}$.

The current into the system from the left lead can be calculated with the system's Green functions as[3]

$$J_L = 2\text{Re}\int \frac{d\omega}{2\pi}\text{Tr}\{[G^<(\omega)\Sigma_L^a(\omega) - G^r(\omega)\Sigma_L^<(\omega)]\sigma_z \otimes \mathbb{I}_4\}. \qquad (7)$$

The matrix $\sigma_z \otimes \mathbb{I}_4$ denotes that the components of the trace corresponding to the second half of the Nambu spinor, $\Psi_\mathbf{r}^\dagger$, are subtracted from the sum rather than added. The less-than, retarded, and advanced Green functions ($G^<, G^r, G^a$) are calculated by using Eqs. 2 and 5 in the non-equilibrium Green Function (NEGF) formalism, which also calculated the density of states profile at every lattice point. We focus on Andreev Bound State (ABS) current, which occurs at energies lower than the superconducting gap and is responsible for current in the DC Josephson Effect. The model is particle-hole symmetric; currents and densities of states are thus equivalent on either side of the Dirac. Small oscillations in the plots, are attributed to DOS confinement from small system size and don't pertain to a physical value. DOS and supercurrent data is low-pass filtered to mitigate noise involved with this confinement.

The simulations use chemical potential as the degree of freedom and are normalized and mapped linearly onto the gate voltage axis of the experimental data so that the Dirac points and bulk band minima align between the model and the sample. Chemical potentials higher than 2.0 eV are beyond the capabilities of the low-energy effective model, hence it is difficult to determine with certainty the nature of critical supercurrent and DOS profiles beyond the ranges shown in Fig. 4.

At low chemical potential, a vanishing surface DOS at the Dirac point results in a minimum of supercurrent that does not completely go to zero. Surface DOS do not fully reach zero at the Dirac point, however, due to the inherent broadening of eigenstates in the Green's function; low-energy eigenstates contribute to the zero-energy spectral weight. This in fact achieves the same result as the experiment, in which surface DOS do not completely vanish at the Dirac point due to disorder-induced inhomogeneous charge puddles, hence the same DOS profiles are achieved due to different underlying principles.

Dimensions are defined in longitudinal length ($\hat{x}$) by transverse width ($\hat{y}$) by depth/thickness ($\hat{z}$), e.g. we use 8×24×8 (1:3 length to width ratio) and 16×8×8 (2:1 length to width ratio) lattice points for the short and long junctions, respectively. Open boundary conditions (BCs) are used in all directions to achieve the appropriate DOS profiles. These dimensions were used to study the effect of junction geometry on critical current; smaller dimensions result in severe finite size confinement, and larger sizes quickly become computationally intractable. The contacts run along the entire width of the device on the top surface ($z = 1$).

To model bulk disorder inherent to the highly doped samples, we add an on-site energy to the system Hamiltonian with magnitude proportional to the severity of disorder. The model corresponds to a system containing a high concentration of $\delta$-function impurities, and allows us to study the dependence of elastic scattering off nonmagnetic impurities in the junction. The on-site energy at each lattice point of the system is randomly generated inside an energy window $W_E$ that satisfies the inequality $-W_E < \varepsilon_w - \varepsilon_o < W_E$, where the on-site energy of the system with (without) disorder is $\varepsilon_{w(o)}$. We investigate two scenarios, one in which only the bulk (at least two lattice points away from any surface) of the junction is disordered, and one in which the entire channel is disordered. We focus on disorder in the long channel geometry as it requires significantly less computation time.